\documentclass[aps,prc,twocolumn,superscriptaddress,showpacs]{revtex4}
\usepackage{graphicx}
\usepackage{sidecap}
\sidecaptionvpos{figure}{c}
\usepackage{multirow}
\usepackage{threeparttable}
\usepackage{tabularx}
\usepackage{hyperref}
\hypersetup{colorlinks,citecolor=blue,filecolor=blue,linkcolor=blue,urlcolor=blue}
\usepackage{color}
\usepackage{dcolumn}

\begin{document}

\title{Ordering of the $\mathbf{0\textit{d}_{5/2}}$ and $\mathbf{1\textit{s}_{1/2}}$ proton levels in light nuclei}
\author{C.~R.~Hoffman}
\email[]{crhoffman@anl.gov}
\affiliation{Physics Division, Argonne National Laboratory, Argonne, Illinois 60439, USA}
\author{B.~P.~Kay}
\email[]{kay@anl.gov}
\affiliation{Physics Division, Argonne National Laboratory, Argonne, Illinois 60439, USA}
\author{J.~P.~Schiffer}
\email[]{schiffer@anl.gov}
\affiliation{Physics Division, Argonne National Laboratory, Argonne, Illinois 60439, USA}

\date{\today}

\begin{abstract}
A survey of the available single-proton data in $A\leq17$ nuclei, along with calculations using a Woods-Saxon potential, show that the ordering of the $0d_{5/2}$ and $1s_{1/2}$ proton orbitals are determined primarily by the proximity of the $s$-state proton energy to the Coulomb barrier. This is analogous to the dependence of the corresponding neutron orbitals in proximity to the neutron threshold, that was previously discussed.
\end{abstract}


\maketitle

\section{Introduction~\label{sec:intro}}
It was recently pointed out~\cite{ref:Hof14} that the spacing between the neutron $0d_{5/2}$ and $1s_{1/2}$ single-particle states is in large part due to the lingering of the $\ell=0$ neutron orbital as it approaches the one-neutron threshold. Such behavior is a consequence of the extended $s$-state wave function and is directly relatable to the neutron halo. Here we show that a similar effect occurs for $unbound$ $\ell=0$ proton orbitals, a fact that we did not fully appreciate at the time Ref.~\cite{ref:Hof14} was published.

The repulsive Coulomb interaction displaces proton states to higher energies relative to their neutron analogs and gives rise to a barrier above the particle threshold. The barrier acts to retard the decay of a proton when it becomes unbound, and as a consequence, proton energies do not show any anomalous behaviors near the proton threshold. However, as $s$-state proton energies in the continuum near the potential barrier height, they \emph{do} show a pattern in energy similar to that of their neutral counterparts~\cite{ref:Hof14}. Therefore, the energy of the proton $1s_{1/2}$ orbital near the potential barrier must effect the ordering of the single-particle levels in much the same way as the neutron orbitals do near the particle threshold. 

In this work, we investigate the behavior of single-proton, 1/2$^+$-5/2$^+$ energies, belonging to the the $1s_{1/2}$ and $0d_{5/2}$ orbitals, through a study of the available proton data for light nuclei. The 5/2$^+$ energies were chosen to reference the 1/2$^+$ energies because the $0d_{5/2}$ is unoccupied, a large amount of data exists, and a straight forward comparison to the available single-neutron data can be made. Woods-Saxon calculations have also been carried out, and they do a good job of reproducing the behavior of $s$-state energies near thresholds.

Weakly bound and unbound single-proton excitations in the $sd$ shell have been discussed on numerous occasions. For instance, they have been investigated in terms of proton halo states~\cite{ref:Ris93,ref:Jen04,ref:Jon04}, Coulomb displacement energies~\cite{ref:Nol69}, and the Thomas-Ehrman shift~\cite{ref:Tho51,ref:Ehr51}. A sub-set of the literature has focused specifically on $s$-state protons or on the relative energies of 1/2$^+$ and 5/2$^+$ states in $sd$-shell nuclei, including Refs.~\cite{ref:She75,ref:For95a,ref:Bar96,ref:Bro96,ref:Oga99a,ref:Aoy98,ref:Aoy00,ref:Yan08} and other papers referred to therein.

\section{Woods-Saxon Calculations\label{sec:wsI}}
Calculations were carried out for $\ell=0$ and 2, $1s_{1/2}$ and $0d_{5/2}$ orbitals to elucidate similarities and differences between the trends in the energies of these states or resonances for protons and neutrons. In this work, the interaction of an individual nucleon with all other nucleons was approximated by an effective potential consisting of a Woods-Saxon potential, and Coulomb plus centrifugal terms as appropriate. For a specific nucleus, the Coulomb barrier energy, $B_V$, was defined as the barrier height of the $\ell=0$ proton potential.

The potential geometry was fixed by the standard parameters. Firstly, the nuclear part of the proton and neutron Woods-Saxon potentials were identical. For the central potential, $r_{0}$=1.25~fm, and $a_0$=0.63~fm, were used, where the radius $R$=$r_0A^{1/3}$~fm. For the spin-orbit potential, $V_{so}$=4.03~MeV, $r_{so0}$=1.1~fm, and $a_{so0}$=0.5~fm, was used, where the radius $R_{so}$=$r_{so0}A^{1/3}$~fm. The strength of the spin-orbit potential was fixed to reproduce the experimental $^{17}$O neutron 1/2$^+$-5/2$^+$ energy difference. The Coulomb potential, affecting only the protons, was calculated for a uniform charge distribution with a radius of $R_C$=1.25$A^{1/3}$~fm.  

\begin{figure}[t]
\centering
\includegraphics[width=1\linewidth]{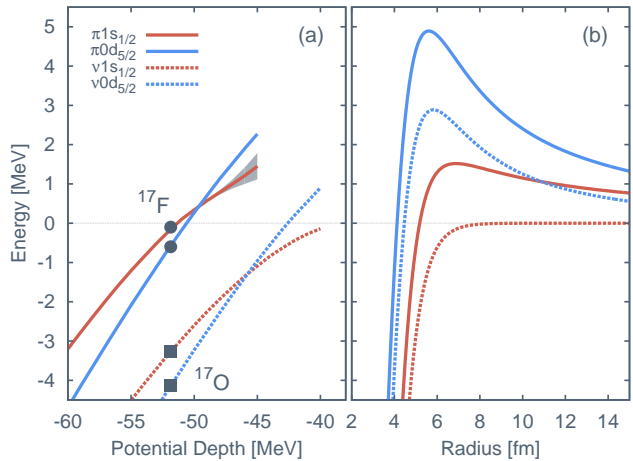}\\
\caption{(Color online) (a) Calculated neutron (dotted) and proton (solid) energies with a Woods-Saxon potential ($^{16}$O plus a nucleon) as a function of its depth for $0d_{5/2}$ (blue) and $1s_{1/2}$ (red) states. The experimental values are shown in $^{17}$O (squares) and $^{17}$F (circles) by the black points at the potential depth required to reproduce the neutron energies. (b) The barrier heights of the effective potentials (Coulomb, plus centrifugal if present) for the same orbitals as in (a).}
\label{fig:fig1}
\end{figure}

The calculated $A=17$ energies are shown as functions of the potential depth ($V$) in Fig.~\ref{fig:fig1}(a). The barrier regions of the effective potentials are displayed in Fig.~\ref{fig:fig1}(b) for this potential. At a fixed depth of $V$=-51.86~MeV, the neutron binding energies in $^{17}$O are reproduced and the calculated energy difference between the proton 1/2$^+$ and 5/2$^+$ levels in $^{17}$F is 0.42~MeV. This is in fair agreement with the experimental value of 0.495 MeV~\cite{ref:Til93}.

%----------------- FIG 1 -------------------------%
%-------------------------------------------------%

%---------------------------FIG 2---------------------%
\begin{figure}[t]
\centering
\includegraphics[width=1\linewidth]{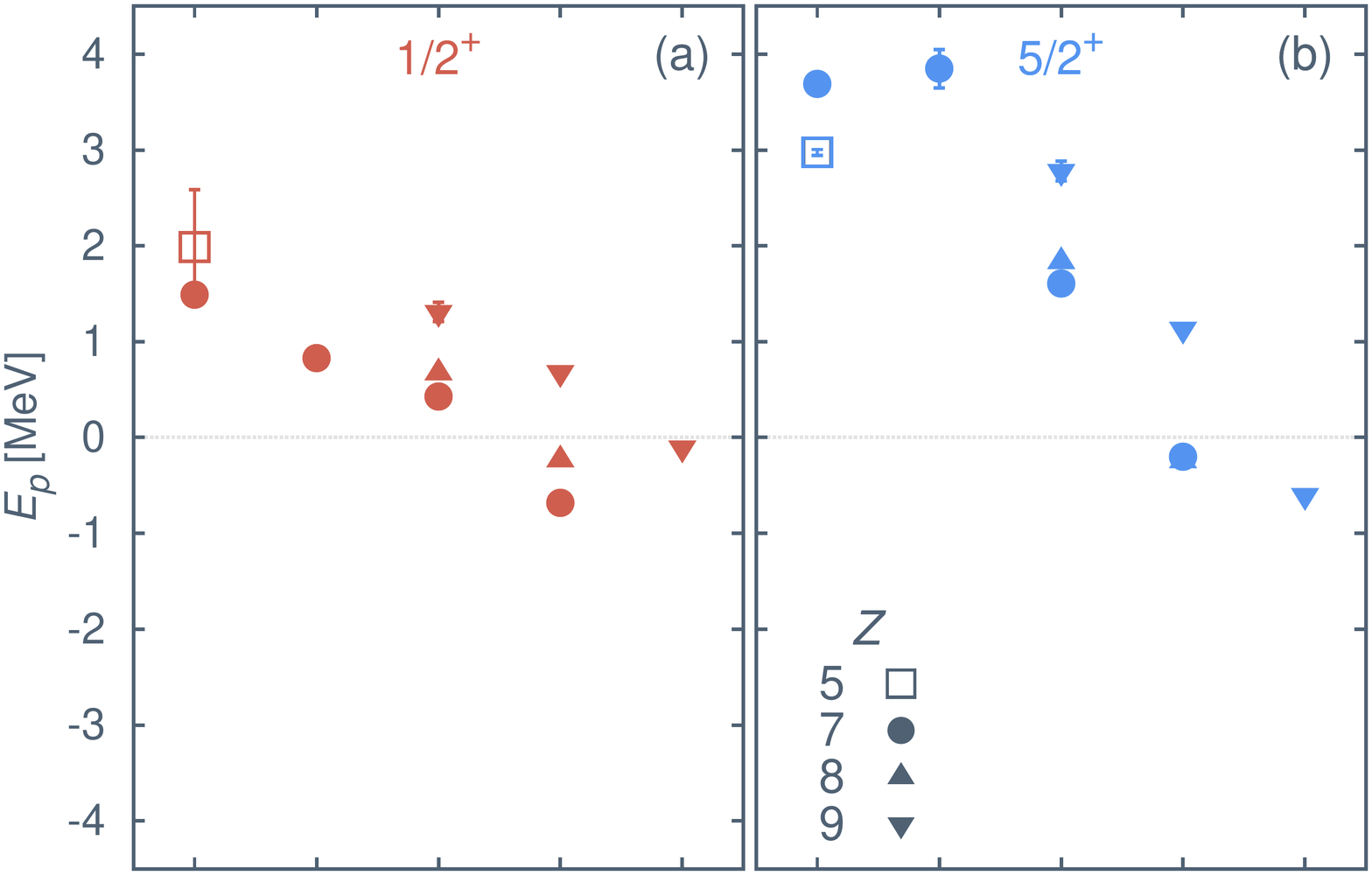}\\
\vspace{-0.55in}
\includegraphics[width=1\linewidth]{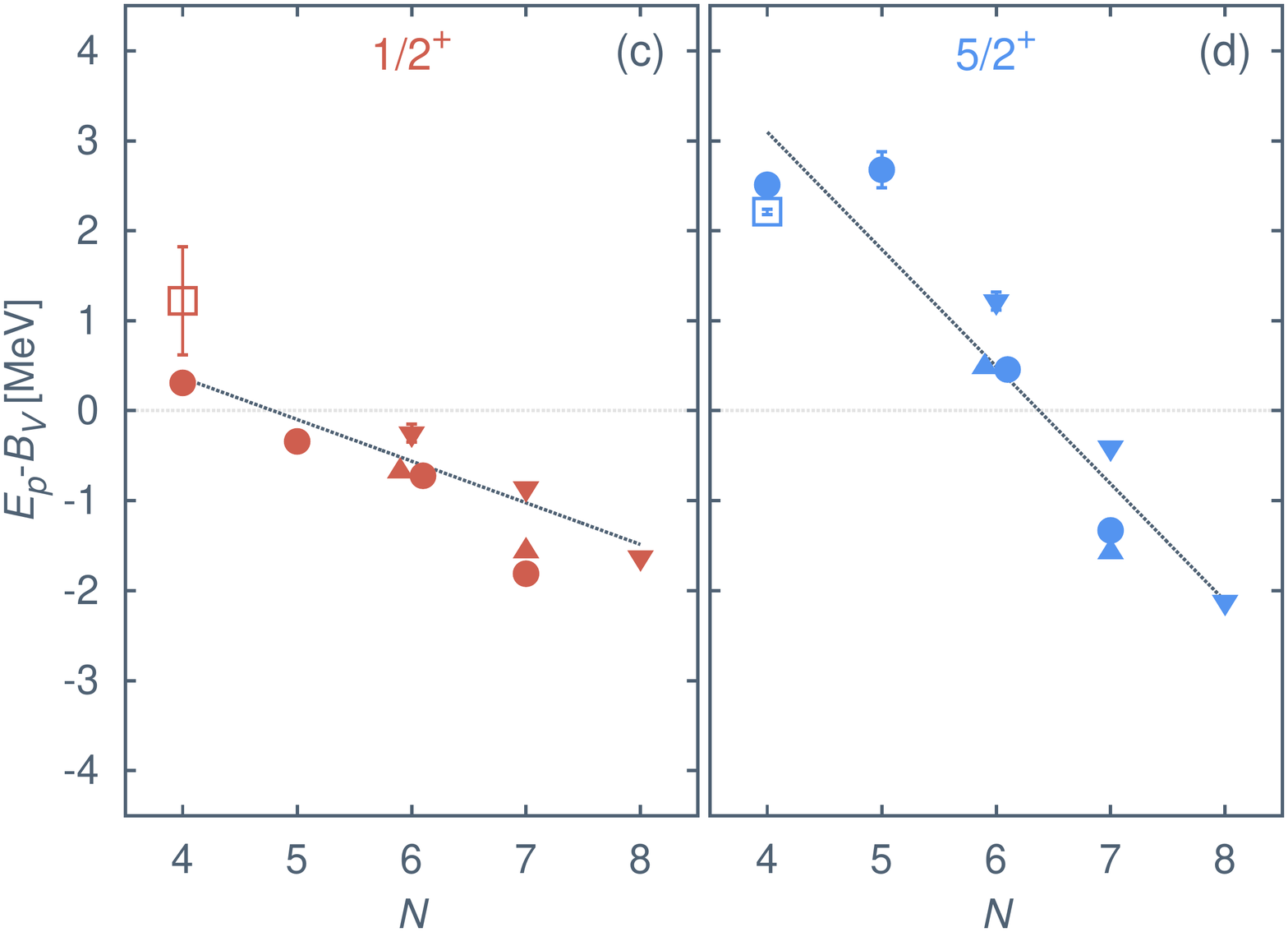}
\caption{(Color online) The data points identify known $A\leq17$, 5/2$^+$ (a) and 1/2$^+$ (b) single-proton energies, $E_{p}$, in $N\leq Z$ nuclei as functions of the neutron number ($N$). (c) and (d) Adjusted single-proton energies, $E_p-B_V$, in which the potential barrier heights have been removed from the energies in (a) and (b). Lines in (c) and (d) are shown to highlight the different data trends. The open symbol for $^{9}$B signifies a lack of convergence as to the 1/2$^+$ resonance energy and width (see text).}
\label{fig:fig2}
\end{figure}
%----------------------------------------------------%

In the continuum, there is little ambiguity in determining resonance energies when they are narrow, but with broader resonances, the procedure needs to be specified. Experimentally, it is often assumed that the resonance energy is at the maximum of the cross section. In more detailed analyses, with the R-matrix formalism~\cite{ref:Lan58}, there are a number of choices to be made. The situation is perhaps least well defined for $\ell=0$ resonances where the phase shift most likley does not pass through $\pi/2$, i.e. $\delta_{max}<\pi/2$. In this work, we converted the calculated Wood-Saxon phase shifts into resonance cross sections, taking the peak value to be the resonance energy. Various prescriptions have been suggested for choosing resonant energies in the literature~\cite{ref:She75,ref:She06}. The range of ambiguity based on a specific choice may be visualized by the grey shaded area in Fig.~\ref{fig:fig1} which shows limits on the $s$-state proton resonance energy based on the calculated resonance width. Overall, conclusions about the mechanism responsible for the sequencing of the $1s_{1/2}$-$0d_{5/2}$ orbitals was not affected by the method used to determine the resonance energies. However, the different prescriptions do lead to variations in the individual energies calculated in Section~\ref{sec:wsII} on the order of the resonance widths.

Crossing of the $\nu1s_{1/2}$ and $\nu0d_{5/2}$ orbitals occurs at around -1.25~MeV in Fig.~\ref{fig:fig1}(a) and is the result of an absence of a barrier for the $\ell=0$ orbital. The radius of the weakly bound $s$-state wave function is extended much further than for cases where there $is$ a barrier: for neutron states with angular momentum or for proton states. Therefore, it is impacted less by changes in the potential, leading to smaller energy changes relative to other orbitals near the particle threshold. The behavior occurs for $s$-state protons in the vicinity of the peak of the potential barrier ($\sim$1.5~MeV for $^{16}$O+$p$), leading to a crossing of the $\pi1s_{1/2}$ and $\pi0d_{5/2}$ orbitals in the continuum at around 0.45~MeV, or $\sim$1.0~MeV below the barrier --- comparable to that for neutrons.

There are clear similarities in the behaviors of $s$-wave neutrons and protons, with both showing slope changes relative to the $\ell=2$ orbital. The commonalities reflect that with the long-range Coulomb repulsion, the same behavior that happens just below the particle threshold for neutrons, also occur just below the top of the potential barrier for protons. The previously realized impact of the potential geometry on the sequence of the single-neutron orbitals~\cite{ref:Hof14} must also have an equally important impact on the single-proton orbitals as well.

%----------------------------TABLE2---------------------------------
\begin{table*}[t]
\caption{\label{tab:tab1} Numerical quantities related to the $1/2^+$ and $5/2^+$ proton single-particle energies. All quantities are in MeV and only those uncertainties that are greater than 10 keV are shown.}
\vspace*{1.5mm}
\newcommand\T{\rule{0pt}{3ex}}
\newcommand \B{\rule[-1.2ex]{0pt}{0pt}}
\begin{ruledtabular}
\begin{tabular}{ldddddddd}
$^{A}Z$ & \multicolumn{1}{c}{$S_p$} & \multicolumn{1}{c}{$E_x(1/2^+)$} & \multicolumn{1}{c}{$E_x(5/2^+)$} & \multicolumn{1}{c}{$E_p(1/2^+)$} & \multicolumn{1}{c}{$E_p(5/2^+)$} & \multicolumn{1}{c}{$\Delta E_{\rm exp}$} & \multicolumn{1}{c}{$\Delta E_{\rm WS}$} & \multicolumn{1}{c}{$B_V$}\\ 
\hline\noalign{\smallskip}
$^{9}$B  & -0.186   & 1.80(60) & 2.79(3) & 1.99(60)  & 2.98(3)  & -0.99(60)& -2.71(22) & 0.77\\
$^{11}$N & -1.49(6) & 0.00     & 2.20(7) & 1.49(6)   & 3.69(7)  & -2.20(9)& -2.37(44)  & 1.18\\
$^{12}$N & 0.601    & 1.43(10) & 4.45(20)& 0.83(10)  & 3.85(20) & -3.02(22)& -1.99(31) & 1.17\\
$^{13}$N & 1.943    & 2.37(5)  & 3.55    & 0.43(5)   & 1.61     & -1.18(5) & -1.07(1)     & 1.15\\
$^{14}$N & 7.551    & 6.87(10) & 7.35(10)& -0.68(10) & -0.20(10)& -0.48(14)& -0.15(12) & 1.13\\
$^{14}$O & 4.630    & 5.30(5)  & 6.46(5) & 0.67(5)   & 1.83(5)  & -1.07(7) & -1.11(8)  & 1.35\\
$^{15}$O & 7.297    & 7.06(5)  & 7.05(5) & -0.24(5)  & -0.25(5) & 0.01(7)  & -0.18(21) & 1.33\\
$^{15}$F & -1.31(10)& 0.00     & 1.47    & 1.31(10)  & 2.78     & -1.47(10) & -1.16(50) & 1.56\\
$^{16}$F & -0.536   & 0.146    & 0.598   & 0.682      & 1.134   & -0.452   & -0.40     & 1.54\\
$^{17}$F & 0.600    & 0.495    & 0.000   & -0.105     & -0.600  & 0.495    & 0.42      & 1.52\\
\end{tabular}
\end{ruledtabular}
\end{table*}
%-----------------------------------------------------------------------------

% Figure----------------------------------------------------%
\begin{figure}[b]
\centering
\includegraphics[width=1\linewidth]{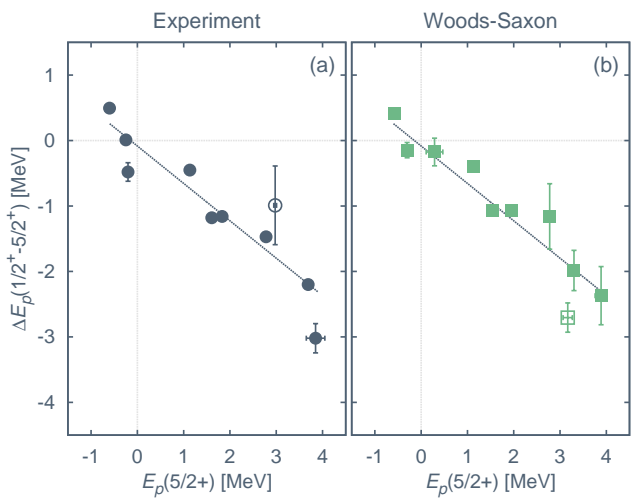}
\caption{(Color online) Experimental (a) and calculated (b) 1/2$^+$ - 5/2$^{+}$ energy differences in proton single-particle excitations, plotted as functions of the 5/2$^+$ proton binding energy.}
\label{fig:fig3}
\end{figure}
%--------------------------------------------------------------------------------------------------------%

\section{Single-proton energy data~\label{sec:data}}
The expected evolution of the proton $1s_{1/2}$ and $0d_{5/2}$ levels based on the Woods-Saxon calculations in Fig.~\ref{fig:fig1}, and the influence of the proximity to the barrier peak on the level sequence, was explored in the available experimental data. In total, ten sets of data for $1/2^+$ and $5/2^+$ single-proton excitations, in nuclei with $Z$ $\geq$ $N$, for $Z=5-9$ and $N=4-8$ were compiled.  Results are shown as a function of neutron number in Figs.~\ref{fig:fig2}(a) and (b) and the numeric values are given in Table~\ref{tab:tab1} of the Appendix. The $^{9}$B data are identified throughout the figures in this work by open symbols due to their dubious nature, in particular, the lack of consensus in the literature as to the energy and width of the 1/2$^+$ resonance, see for example Ref.~\cite{ref:For13b} and references therein.

Single-proton energies relative to the proton-particle threshold are $E_p(J^{\pi})=E_x(J^{\pi})-S_{p}$, where $S_{p}$ is the one-proton separation energy so that positive values of $E_{p}(J^{\pi})$ are particle unbound. $E_{x}(J^{\pi})$ is the excitation energy of the state (or the centroid of states) belonging to a specific orbital, $J^{\pi}$=1/2$^+$ or 5/2$^+$.

The trends of the 1/2$^+$ and $5/2^+$ excitations (Fig.~\ref{fig:fig2}) qualitatively resemble those from the Woods-Saxon calculations of Fig.~\ref{fig:fig1}(a). The 1/2$^+$ energies change by $\sim$2~MeV from $N=4$ to 8 and trend toward the region of the potential barrier heights $<2$~MeV for $Z<9$. The 5/2$^+$ energies change by $\sim$5~MeV over the same range of neutron numbers. The scatter of multiple data points at a single $N$ value is caused, in part, by the differing $Z$ values. This is further evidenced by the systematic shifts, although each having similar slopes, of the different $Z$ cores, e.g., the $Z=9$ points are generally higher in energy than all other data.

Because we are specifically concerned with the proton $s$-state behavior relative to the potential barrier height, $B_V$, we subtract it from both sets of proton energies and plot them as a function of neutron number [Fig.~\ref{fig:fig2}(c) and (d)]. As mentioned previously, the barrier height energies are estimated from the effective Woods-Saxon potentials (Fig.~\ref{fig:fig1}) for each individual case. Removal of the barrier energy allows for a direct comparison of the proton data to the neutron data of our previous work~\cite{ref:Hof14}. After the barrier subtraction, the 1/2$^+$ and 5/2$^+$ energies still show distinct trends, as emphasized by the guidelines in Fig.~\ref{fig:fig2}(c) and (d), respectively. Also, the lingering behavior in the proton $s$-state data now occurs around $\sim$0~MeV, similar to the neutron data.

%------------------------------------------------%
\begin{figure}[t]
\centering
\includegraphics[width=1\linewidth]{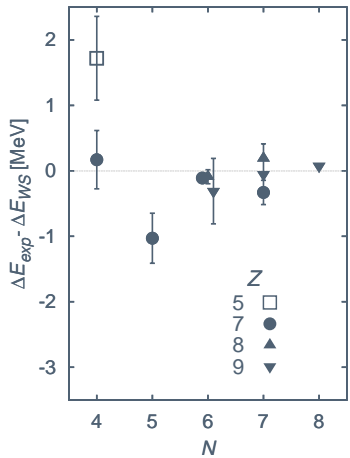}
\caption{Difference between the experimental [Fig.~\ref{fig:fig3}(a)], and calculated [Fig.~\ref{fig:fig3}(b)], $\Delta E_p$(1/2$^+$-5/2$^+$) energy differences.}
\label{fig:fig4}
\end{figure}
%------------------------------------------------------%

%--------------------------------------------------------------------%
\begin{figure*}[tb]
\centering
\includegraphics[width=0.5\linewidth]{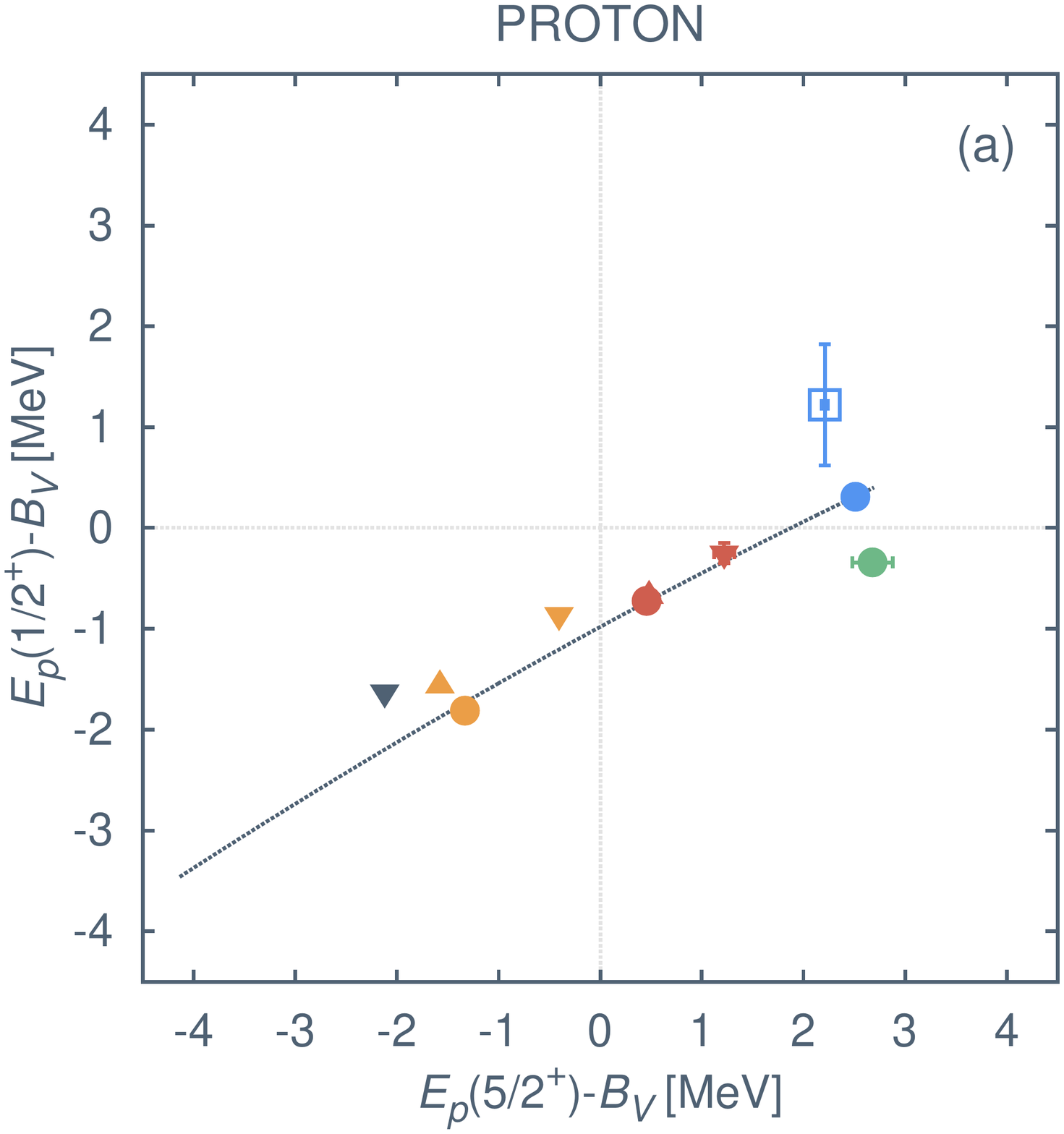}
\hspace{-0.5in}
\includegraphics[width=0.5\linewidth]{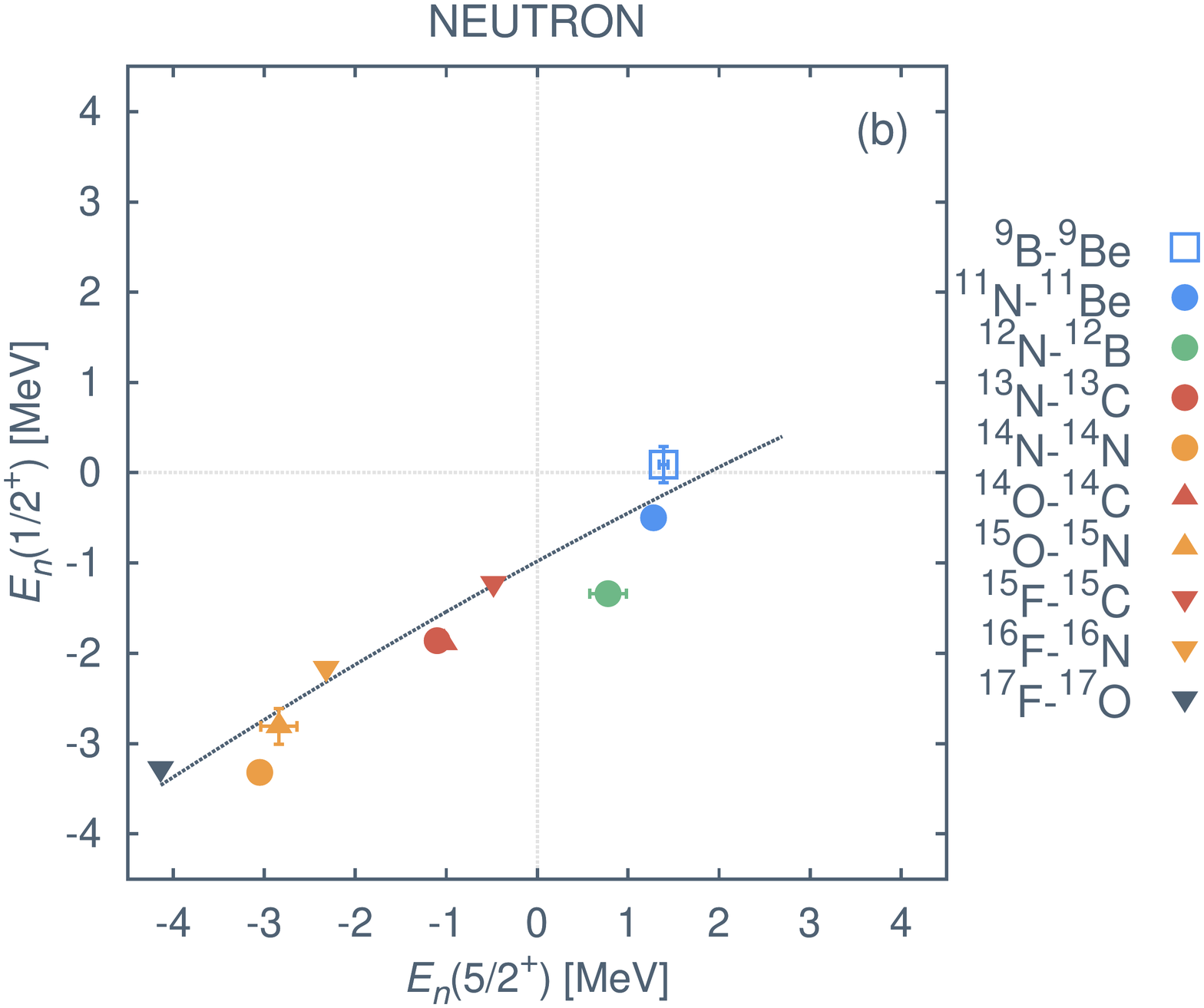}\\
\caption{(Color online) (a) The 1/2$^+$ barrier height adjusted proton excitations plotted against the adjusted 5/2$^+$ excitations. (b) The same plot as in (a) but for the neutron excitations in the mirror nuclei. The solid line in each figure is identical and meant to guide the eye.}
\label{fig:fig5}
\end{figure*}
%--------------------------------------------------------------%

Changes in the relative energies between the $\pi1s_{1/2}$ and $\pi0d_{5/2}$ orbitals are shown in Fig.~\ref{fig:fig3}(a) by the energy difference $\Delta E_p(1/2^+-5/2^+)=E_{p}(1/2^+)-E_{p}(5/2^+)$ as a function of the proton 5/2$^+$ binding energy, $E_{p}(5/2^+)$. Crossing of the $1/2^+$ and $5/2^+$ data occurs around $E_p\approx0$~MeV, as expected from the $A=17$ calculations but at slightly lower energy [Fig.~\ref{fig:fig1}(a)].

Additional information on the $E_x(J^{\pi})$ values used in various nuclei, such as the specific choices of energies and the sources of uncertainties, is provided in the Appendix. In nuclei with odd-$Z$, odd-$A$ and even-even cores ($^{9}$B, $^{11,13}$N, and $^{15,17}$F), the $E_{x}(J^{\pi})$ are the energies of the lowest-lying single states with $J^{\pi}$=1/2$^+$ or 5/2$^+$. These states have been, to a large degree, identified to be single-particle in nature by having large cross sections in transfer reactions, and spectroscopic overlaps consistent with single-particle states, or by having resonance widths approaching single-particle estimates. In odd-$Z$, odd-$N$, $Z>N$ nuclei ($^{12}$N and $^{16}$F), $E_{x}(J^{\pi})$ energy centroids are determined from the ($2J+1$)$C^2S$-weighted averages of the multiplet of states created with the odd nucleon. For the two cases of $Z=8$ ($^{14}$O and $^{15}$O), the $E_{x}(J^{\pi})$ energy centroids are also determined from the ($2J+1$)$C^2S$-weighted averages of the multiplet of states. In these cases, all observed $\ell=0$ or 2 strength below an excitation energy of $E_{x}\sim8.5$~MeV, was assumed to be part of the 1/2$^+$ or 5/2$^+$ centroids. In $^{14}$N, $E_{x}(J^{\pi})=1/2(E_{T=0}+E_{T=1})$, where $E_{T=0,1}$ are the ($2J+1$)$C^2S$-weighted energy centroids for isospin 0 and 1, respectively.

\section{Calculated Proton Energies~\label{sec:wsII}}
$1s_{1/2}$ and $0d_{5/2}$ proton energies were calculated with a Woods-Saxon potential for each of the ten nuclei for which there were proton excitation data. The parameters detailed in Sec.~\ref{sec:wsI} were used. For each nucleus, the potential depth of the Woods-Saxon was fixed so as to reproduce the experimental \emph{neutron} $5/2^+$ energy centroid of its corresponding mirror nucleus (Table~II of Ref.~\cite{ref:Hof14}). For example, the potential depth in $^{17}$O was fixed to $V$=-51.86~MeV, and this was used in the calculation of \emph{both} states in $^{17}$F.

The calculated $\pi1s_{1/2}$-$\pi0d_{5/2}$ energy differences are shown alongside the experimental data in Fig.~\ref{fig:fig3}(b) and given in Table~\ref{tab:tab1}. There are systematic uncertainties in the calculations from the choice of the Woods-Saxon parameters and uncertainties in the experimental neutron 5/2$^+$ energies used to constrain them. There is an additional systematic uncertainty caused by ambiguities in determining a proton $1s_{1/2}$ resonance energy for broad states (discussed in Section~\ref{sec:wsI}).

The fact that the Coulomb energy differences cannot be reproduced exactly is well known, and the Nolen-Schiffer Anomaly~\cite{ref:Nol69}, a 10\% effect caused by charge-symmetry violation in the $NN$ interaction, is well established. However, here we adhere to discussions of the relative energies between the two orbitals where this effect would tend to cancel. In addition, through the use of an effective potential to reproduce the change in orbital energies, we do not attempt to explicitly explore the role of correlations.

The change in the relative energies of the $\pi1s_{1/2}$-$\pi0d_{5/2}$ orbitals, as a function of binding energy, is reproduced by the Woods-Saxon calculations when the proper potential geometry is used. The residuals between the experimental 1/2$^+$-5/2$^+$ energy differences and the calculated $1s_{1/2}$-$0d_{5/2}$ energy differences are shown in Fig.~\ref{fig:fig4} as a function of neutron number. The calculations accounted for the $\sim4$~MeV relative energy change that occurred between the two orbitals.

Some variation of $\Delta E_{exp}-\Delta E_{WS}$ away from zero is expected due to the monopole component of the tensor force. This was observed in the neutron residuals (Fig. 5 of Ref.~\cite{ref:Hof14}) and the effect should be the same for these mirror states. While not acting on the $s$-state proton orbital, the tensor force will influence the 5/2$^+$ excitations as the neutron number ($p$-shell occupancy) changes. As protons are removed from the $\nu0p_{1/2}$ orbital, the action of the tensor force would cause a decrease in the residuals with a minimum at $N=6$ (assuming a sharp separation between $0p_{1/2}$ or $0p_{3/2}$ orbits filling, above or below $N=6$). The present data are consistent with the expected effect of the tensor interaction, however, the error bars are far too large, and the data too scattered, to ascertain anything quantitative, especially below $N=6$.

%--------------------------------------------------------------------%
\begin{figure}[tb]
\centering
\includegraphics[width=1\linewidth]{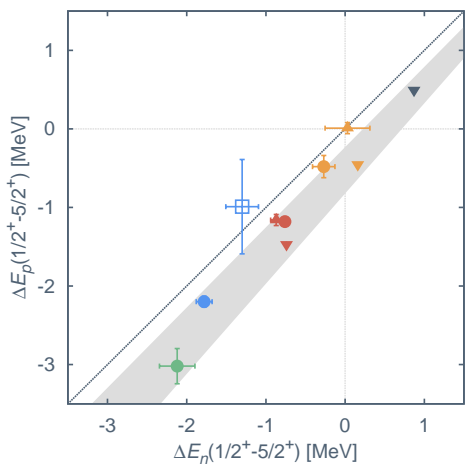}\\
\caption{(Color online) The 1/2$^+$-5/2$^+$ energy difference between the proton states is plotted against the same energy difference for the mirror neutron states. The data points are labeled in the same way as in Fig.~\ref{fig:fig5}. The diagonal line has a slope equal to one representing a 1:1 correlation between the complementary data sets. The grey band identifies the region covered by Woods-Saxon calculations of the energy differences for the various nuclei.}
\label{fig:fig6}
\end{figure}
%--------------------------------------------------------------%

\section{proton and neutron data~\label{sec:pn}}
The trends of the $\ell=0$ proton and neutron data are intertwined by common approaches to their respective barriers, the Coulomb for protons and the particle threshold for neutrons. Subtracting the potential barrier height from the proton data facilitates a comparison with the neutron data, however, the comparison cannot be perfect since the Coulomb potential, albeit long range, $does$ change with radius. The 1/2$^+$ and 5/2$^+$ proton excitation data with barrier corrections, are plotted against one-another in Fig.~\ref{fig:fig5}(a) and the neutron excitation data for corresponding mirror nuclei are plotted against one-another in Fig.~\ref{fig:fig5}(b). A line, common to both figures, highlights similarities in the data and guides the eye. Agreement between the data in the two plots gives credence to the common, near-threshold behavior of all $s$-states.

Small shifts in the energies of the proton and neutron data, above and below the common line, respectively, are noticed in Fig.~\ref{fig:fig5}(a). While there are uncertainties in the data, the Thomas-Ehrman shift~\cite{ref:Tho51,ref:Ehr51} must be considered when comparing loosely-bound (or unbound) proton and neutron states. The shift depends on the difference in radial distributions and the root-mean-square radii of $s$-states will be larger than those of $d$-states. This is included in Woods-Saxon calculations and so it does not affect a comparison between data and calculations. However, it is $not$ included when neutron data are compared to proton data.

The proton $\Delta E_p(1/2^+-5/2^+)$ energy is plotted against the neutron $\Delta E_n(1/2^+-5/2^+)$ energy in Fig.~\ref{fig:fig6} for mirror nuclei (identified by the key in Fig.~\ref{fig:fig5}). The diagonal line in Fig.~\ref{fig:fig6} represents the proton and neutron energy differences being the same. Note how the only data point north-west of the line is from the pesky $A$=9 mirrors (including the controversial $^9$B 1/2$^+$ energy), although the error bar is large.

To the extent that the data points in Fig.~\ref{fig:fig6} are on a line parallel to the diagonal, it implies that the Thomas-Ehrman shift has an approximately fixed value in the nuclei studied in this work. We note, that in general, the more loosely bound states are also of lower $Z$, and both $Z$ and the binding energy will have an influence on the Thomas-Ehrman shift. In addition, the Woods-Saxon calculations of Section~\ref{sec:wsII} qualitatively support this, because they lie on the proper side of the diagonal and vary little as shown by the grey band in Fig.~\ref{fig:fig6}. The width of the band accounts for variations in potential parameters and in the extraction of the resonance energies. This is expected due to the good agreement with the data as shown in Fig.~\ref{fig:fig4} for protons and Fig.~5 of Ref.~\cite{ref:Hof14} for neutrons.

\section{Conclusions~\label{sec:sum}}
In this work, we point out that the behavior observed for $s$-state neutrons near the particle threshold, extends to $s$-state protons in the continuum, as well, when the proton energies are near the peak of the effective potential barrier. Therefore, changes in the relative proton energies of the $1s_{1/2}$ and $0d_{5/2}$ orbitals are primarily due to the proximity of the $s$-state proton to the Coulomb barrier, making the impact on the nuclear level ordering due to $s$-state behavior near threshold a general feature of nuclei.

Conclusions were drawn from completing a systematic survey of ten sets of 1/2$^+$ and 5/2$^+$ proton centroid data in $A\leq$17 nuclei, by comparing these data to Woods-Saxon calculations, and by comparing the proton data to available mirror neutron excitation data. One-body potential calculations with a Woods-Saxon potential, using fixed parameters based on known \emph{neutron} data, reproduced the relative proton 1/2$^+$-5/2$^+$ energy-difference trend and isolated the importance of the potential geometry. In addition, after adjusting the proton data by their potential barrier heights, the binding energy relation between 1/2$^+$-5/2$^+$ excitations was found to be the same as that for the neutron data. Such consistency over the nuclei that were surveyed identifies a commonality within the two data sets of the $s$-state behavior near a barrier while also eluding to a near-constant Thomas-Ehrman shift in this region.

\section*{Acknowledgments~\label{sec:ack}}
The authors thank a number of colleagues for helpful discussions. We also appreciate the helpful comments on the manuscript from P.~W.~Zhao, S. Bottoni, and D. Santiago-Gonzalez. This work was supported by the U.S. Department of Energy, Office of Nuclear Physics, under Contract No. DE-AC02-06CH11357.

\section*{Appendix~\label{sec:app}}

The data used in the present work are given in Table~\ref{tab:tab1}. Additional information on the extraction of the single-proton energies for the states of interest is included below. As pointed out in the text, resonance energies from the Woods-Saxon calculations were determined at the energy in which the cross section, $\sigma$, reached its maximum value. Single-particle widths, $\Gamma_{sp}$, were determined by the calculated Woods-Saxon resonance width at 1/2 its maximum after the potential depth had been adjusted so that $\sigma_{max}$ was at the empirical resonance energy. This procedure can be plagued by similar effects as the resonance energy determinations when the widths become large. No data were reanalyzed, but the calculated widths were used to check for the single-particle characteristics of a particular state or states.

\textbf{$^{9}$B:}
$S_p$ was taken from the 2012 Atomic Mass Evaluations (AME)~\cite{ref:Aud12}. The energy of the 1/2$^+$ resonance has been infamously elusive for many years as pointed out in the text. Here we adopted a value of 1.80(60)~MeV based on the most recent work~\cite{ref:Whe15} and those therein. In addition, the reported width of this state [$\Gamma=0.65(13)$~MeV]~\cite{ref:Whe15} is well below the expected single-particle width [$\Gamma_{sp}\gtrsim4$~MeV]. A large uncertainty of 0.6~MeV has been arbitrarily assumed to cover the large possible range of excitation energy values and the uncertainty in the single-proton nature of the state. The 5/2$^+$ excitation energy was taken from Table 9.13 of the latest compilation~\cite{ref:Til04}. The width of this state [$\Gamma=0.55(4)$~MeV]~\cite{ref:Til04} is consistent with a single-proton state [$\Gamma_{sp}\approx0.6$~MeV] so there is no additional uncertainty.

\textbf{$^{11}$N:}
$S_p$ was taken from Table 11.45 of Ref.~\cite{ref:Kel12}. The excitation energy of the 1/2$^+$ state was taken from Table 11.45 of Ref.~\cite{ref:Kel12}. The width of this state [$\Gamma=0.83(3)$~MeV]~\cite{ref:Kel12} is well below the expected single-particle width [$\Gamma_{sp}\gtrsim4$~MeV]. An arbitrary uncertainty of 100~keV on the excitation energy is assumed due to uncertainty in the single-proton nature of the state. The 5/2$^+$ excitation energy was taken from Table 11.45 of Ref.~\cite{ref:Kel12}. The width of this state [$\Gamma=0.54(4)$~MeV]~\cite{ref:Kel12} is consistent with a single-proton state [$\Gamma_{sp}\approx0.8$~MeV] so there is no additional uncertainty.

\textbf{$^{12}$N:}
$S_p$ was taken from the 2012 AME~\cite{ref:Aud12}. The odd neutron splits the proton $1s_{1/2}$ orbital into a $2^-$ and $1^-$ multiplet of states. The excitation energies of these states, 1.20 and 1.80~MeV, respectively, were taken from Refs.~\cite{ref:Per06,ref:Chi15}. The 1/2$^+$ centroid excitation energy was extracted from these two states by a $(2J+1)C^2S$ weighted averaging. $C^2S=1$ was assumed for both states. The width of the $2^-$ state [$\Gamma=0.12(1)$~MeV]~\cite{ref:Chi15} is consistent with a single-proton state [$\Gamma_{sp}\approx0.14$~MeV] but the $1^-$ states width [$\Gamma=0.75(25)$~MeV]~\cite{ref:Chi15} is below the expected single-particle width [$\Gamma_{sp}\approx2.0$~MeV]. An arbitrary uncertainty of 100~keV on the 1/2$^+$ centroid energy is assumed due to the uncertainty in the single-proton nature of the 1$^-$ state and the $C^2S=1$ assumption.

The odd neutron splits the proton $0d_{5/2}$ orbital into a $1^-$, $2^-$, $4^-$, and $3^-$ multiplet of states. The excitation energies of these states, 3.43~MeV, 3.98~MeV, 4.34, and 5.35~MeV, respectively, were taken from Refs.~\cite{ref:Per06,ref:Chi15}. There are uncertainties in these assignments, in particular with the $J^{\pi}=3^-$ level. The 5/2$^+$ centroid excitation energy was extracted from these four states by a $(2J+1)C^2S$ weighted averaging. $C^2S=1$ was assumed for all states. The widths of these states were not investigated since questions regarding their single-proton nature already existed. An uncertainty of 200~keV on the 5/2$^+$ centroid energy is assumed due to the dubious spin-assignments of some of the states and the $C^2S=1$ assumption.

\textbf{$^{13}$N:}
S$_{p}$ was taken from the 2012 AME~\cite{ref:Aud12}. The 1/2$^+$ excitation energy was taken from Table 13.14 of Ref.~\cite{ref:Ajz91}. The width of this state [$\Gamma=0.032(1)$~MeV]~\cite{ref:Ajz91} is consistent with a single-proton state [$\Gamma_{sp}\approx0.06$~MeV], however, small spectroscopic factors were found in Ref.~\cite{ref:Pet80}. Hence, an uncertainty of 50 keV on the excitation energy is assumed. The 5/2$^+$ excitation energy was taken from Table 13.14 of Ref.~\cite{ref:Ajz91} also. The width of this state [$\Gamma=0.047(7)$~MeV]~\cite{ref:Ajz91} is consistent with a single-proton state [$\Gamma_{sp}\approx0.06$~MeV] so there is no additional uncertainty.

\textbf{$^{14}$N:}
$S_p$ was taken from the 2012 AME~\cite{ref:Aud12}. The 1/2$^+$ energy comes from $E=1/2(E_{T=0}+E_{T=1})$ where $E_{T=0,1}$ corresponds to the $2J+1$ weighted centroids ($C^2S=1$) of the two strongest $T=0,1$ 0$^-$ and 1$^-$ states (4.92, 5.69, 8.06, and 8.78~MeV) in $^{14}$N identified in the $^{13}$C($^{3}$He,$d$) reaction and listed in Table~14.18 of Ref.~\cite{ref:Ajz91}. The same procedure was used for the 5/2$^+$ centroid using the four 2$^-$ and 3$^-$ states (5.11, 5.83, 8.91, and 9.51~MeV). A correction factor of one-half of the energy difference between the states in $^{13}$C-$^{13}$N was applied for the neutron states but no such correction was applied here for the proton states. An uncertainty of 100 keV is assumed due to possible fragmentation.

\textbf{$^{14}$O:}
$S_p$ was taken from the 2012 AME~\cite{ref:Aud12}. The proton $1s_{1/2}$ orbital is split into a 1$^-$ and $0^-$ multiplet due to the remaining odd $0p_{1/2}$ proton. The excitation energies of these states, 5.16 and 5.71~MeV, respectively, were taken from Table 1 of Ref.~\cite{ref:Ter07}. The 1/2$^+$ centroid excitation energy was extracted from these two states by a $(2J+1)C^2S$ weighted averaging. $C^2S=1$ was assumed for both states. The width of both the $1^-$ state [$\Gamma=0.042(4)$~MeV] and the $0^-$ state [$\Gamma=0.40(10)$~MeV]~\cite{ref:Ter07} are consistent with single-proton states [$\Gamma_{sp}\approx0.05$~MeV and $\Gamma_{sp}\approx0.55$~MeV, respectively]. An uncertainty of 50~keV is assumed on the 1/2$^{+}$ centroid energy due to the $C^2S=1$ assumption.

The proton $0d_{5/2}$ orbital is split into a 3$^-$ and $2^-$ multiplet due to the remaining odd $0p_{1/2}$ proton. The excitation energies of these states, 6.23 and 6.77~MeV, respectively, were taken from Table 1 of Ref.~\cite{ref:Ter07}. The 5/2$^+$ centroid excitation energy was extracted from these two states by a $(2J+1)C^2S$ weighted averaging. $C^2S=1$ was assumed for both states. The width of both the $3^-$ state [$\Gamma=0.042(2)$~MeV] and the $2^-$ state [$\Gamma=0.090(5)$~MeV]~\cite{ref:Ter07} are consistent with single-proton states [$\Gamma_{sp}\approx0.05$~MeV and $\Gamma_{sp}\approx0.13$~MeV, respectively]. An uncertainty of 50~keV is assumed on the 5/2$^{+}$ centroid energy due to the $C^2S=1$ assumption.

\textbf{$^{15}$O:}
$S_p$ was taken from the 2012 AME~\cite{ref:Aud12}. The proton $1s_{1/2}$ orbital is split into various states having $J^{\pi}$=1/2$^{+}-3/2^{+}$ due to the odd $0p_{1/2}$ neutron and extra unpaired $0p_{1/2}$ proton. The excitation energy of these states were taken from Table 15.16 of Ref.~\cite{ref:Ajz91}. The 1/2$^+$ centroid included states at 5.183 [1/2$^+$], 6.793 [3/2$^+$], and 7.557~MeV [1/2$^+$]. The 1/2$^+$ centroid excitation energy was extracted from these states by a $(2J+1)C^2S$ weighted averaging of all $\ell$=0 strength. The $C^2S$ values was taken from the $^{14}$N($^{3}$He,$d$) data given in the ``Present results'' column of Table 2 in Ref.~\cite{ref:Alf69}. An uncertainty of 50 keV is assumed on the 1/2$^{+}$ energy centroid due to the large fragmentation of the $\ell$=0 strength. 

The proton $0d_{5/2}$ orbital is split into various states having $J^{\pi}$=3/2$^{+}-7/2^{+}$ due to the odd $0p_{1/2}$ neutron and extra unpaired $0p_{1/2}$ proton. The excitation energy of these states were taken from Table 15.16 of Ref.~\cite{ref:Ajz91} and for the 5/2$^+$ centroid included states at 5.241 [5/2$^+$], 6.859 [5/2$^+$], 7.276 [7/2$^+$], and 8.284~MeV [3/2$^+$]. The 5/2$^+$ centroid excitation energy was extracted from these states by a $(2J+1)C^2S$ weighted averaging of all $\ell$=2 strength. The $C^2S$ values was taken from the $^{14}$N($^{3}$He,$d$) data given in the ``Present results'' column of Table 2 in Ref.~\cite{ref:Alf69}. An uncertainty of 50 keV is assumed on the 5/2$^{+}$ energy centroid due to the large fragmentation of the $\ell$=2 strength and the assumption that all $\ell=2$ strength belongs to the $0d_{5/2}$ orbital (as opposed to the higher-lying $0d_{3/2}$ orbital).

\textbf{$^{15}$F:}
S$_{p}$ was taken from the energy of the lowest-lying resonance in Ref.~\cite{ref:Ste14}. The energy of this resonance varies in the literature~\cite{ref:Gol04,ref:Guo05,ref:Muk08,ref:Muk09,ref:Muk10,ref:Ste14} so an uncertainty of 100~keV on the proton separation energy is assumed. The excitation energy of the 1/2$^{+}$ state was adopted from the work of Ref.~\cite{ref:Ste14}. The width of this state [$\Gamma=0.85(15)$~MeV] is consistent with a single-proton state [$\Gamma_{sp}\approx0.9$~MeV] so there are no additional uncertainties. The excitation energy of the 5/2$^{+}$ state was adopted from the work of Ref.~\cite{ref:Ste14}. The width of this state [$\Gamma=0.31(1)$~MeV] is consistent with a single-proton state [$\Gamma_{sp}\approx0.3$~MeV] so there are no additional uncertainties.

\textbf{$^{16}$F:}
$S_p$ was taken from the 2012 AME~\cite{ref:Aud12}. The proton $1s_{1/2}$ orbital is split into a 0$^-$ and 1$^-$ multiplet due to the odd $0p_{1/2}$ neutron. The excitation energies of these states, 0.0 and 0.194~MeV, respectively, were taken from the ``New recommended'' column of Table 1 in Ref.~\cite{ref:Ste14}. The 1/2$^+$ centroid excitation energy was extracted from these two states by a $(2J+1)C^2S$ weighted averaging. $C^2S=1$ was assumed for both states. The widths of both states are consistent with single-proton states, Table 1 of Ref.~\cite{ref:Ste14}, so there are no additional uncertainties. The proton $0d_{5/2}$ orbital is split into a 2$^-$ and 3$^-$ multiplet due to the odd $0p_{1/2}$ neutron. The excitation energies of these states, 0.425 and 0.721~MeV, respectively, were taken from the ``New recommended'' column of Table 1 in Ref.~\cite{ref:Ste14}. The 5/2$^+$ centroid excitation energy was extracted from these two states by a $(2J+1)C^2S$ weighted averaging. $C^2S=1$ was assumed for both states. The widths of both states are consistent with single-proton states, Table 1 of Ref.~\cite{ref:Ste14}, so there are no additional uncertainties.

\textbf{$^{17}$F:}
$S_p$ was taken from the 2012 AME~\cite{ref:Aud12}. The excitation energy of the 1/2$^+$ state was taken from Table 17.23 of Ref.~\cite{ref:Til93}. This state was found to have large cross sections and spectroscopic factors in proton transfer measurements, e.g., see Refs.~\cite{ref:Oli69,ref:For75a,ref:Ajz82}, which is indicative of a single-proton state. The excitation energy of the 5/2$^+$ state was taken from Table 17.23 of Ref.~\cite{ref:Til93}. This state was also found to have large cross sections and spectroscopic factors in proton transfer measurements, e.g., see Refs.~\cite{ref:Oli69,ref:For75a,ref:Ajz82}, which is indicative of a single-proton state.

\bibliography{prc}

\end{document}